\documentclass[aps,prc,twocolumn,nofootinbib,showpacs,superscriptaddress,groupedaddress]{revtex4-1}
\usepackage{amsmath,amssymb,amsbsy,bm}
\usepackage{graphicx}
\usepackage{comment}
\usepackage{float}
\usepackage[colorlinks=true,linkcolor=blue,citecolor=blue,urlcolor=blue]{hyperref}

\begin{document}

\title{Determining the Diffusivity for Light Quarks from Experiment}
\author{Scott Pratt}
\affiliation{Department of Physics and Astronomy and National Superconducting Cyclotron Laboratory\\
Michigan State University, East Lansing, MI 48824~~USA}
\author{Christopher Plumberg}
\email{christopher.plumberg@gmail.com}
\thanks{ORCID: https://orcid.org/0000-0001-6678-3966}
\affiliation{Department of Astronomy and Theoretical Physics\\
Lund University, S\"{o}lvegatan 14A, SE-223 62, Lund, Sweden}
\date{\today}


\begin{abstract}
Charge balance functions reflect the evolution of charged pair correlations throughout the stages of pair production, dynamical diffusion, and hadronization in heavy-ion collisions. Microscopic modeling of these correlations in the full collision volume shows that the balance functions are sensitive to the diffusivity of light quarks when studied as functions of relative azimuthal angle. By restricting our analysis to $K^+K^-$ and $p\bar{p}$ pairs, we find that the diffusivity of light quarks, a fundamental property not currently well understood, can be constrained by experimental measurement.
\end{abstract}

\maketitle

A principal goal of relativistic heavy ion physics is to determine bulk properties of the quark-gluon plasma (QGP), i.e. matter where the density exceeds the point at which individual hadrons can be defined. Interesting properties include the equation of state, charge susceptibility, quark-antiquark condensate, viscosity, diffusivity and jet-energy loss. Several of these properties can be reliably extracted from lattice gauge theory, but even in those cases it is important to extract the properties from experiment to test whether the idealization of a locally equilibrated QGP has indeed been realized in the collision. Careful comparisons of experiment to theoretical models have so far constrained the equation of state \cite{Pratt:2015zsa}, charge susceptibility \cite{Pratt:2015jsa}, viscosity \cite{Pratt:2015zsa,Bernhard:2016tnd,Bernhard:2015hxa}, jet-energy loss \cite{Burke:2013yra,He:2018gks} and the diffusivity for heavy quarks \cite{Xu:2017obm}. Here, we describe how the diffusivity for light quarks can be added to this list.

For temperatures near or above 200 MeV, a range explored in heavy-ion collisions at the Relativistic Heavy Ion Collider (RHIC) and by the LHC at CERN, the charge susceptibility of light quarks is consistent with a picture which treats the light quarks (up, down, and strange) as well-defined quasi-particles, in accordance with lattice calculations \cite{Borsanyi:2011sw,Bellwied:2015lba}. This result is rather surprising given that the medium is strongly interacting. The shear viscosity and diffusivity represent measures of how strongly the matter interacts. Because there are three charges, or flavors, the diffusivity, $D_{ab}$, is a three-by-three matrix,
\begin{eqnarray}
\bm{j}_a&=&-D_{ab}\nabla \delta\rho_b,
\end{eqnarray}
where $\rho_a$ and $\bm{j}_a$ are the charge and current densities in the rest frame of the fluid. As the medium is heated above the hadron/QGP transition temperature ($\sim 160$ MeV), and up, down and strange quarks become good quasi-particles, the three flavors should behave similarly for zero chemical potential, and the matrix should become proportional to the unit matrix, $D_{ab}\approx D\delta_{ab}$.\footnote{Note that the indices $a$,$b$ range over the light quark basis $(u,d,s)$.  A basis of conserved charges $(B,Q,S)$ would not yield a diagonal diffusivity matrix in the QGP phase \cite{Greif:2017byw}.} Such conditions are realized in the mid-rapidity regions for top RHIC and LHC beam energies. Theoretical calculations of the diffusivity are typically based on the Kubo relation for the conductivity tensor, which translates into the diffusivity,
\begin{eqnarray}
\bm{j}_a&=&-\sigma_{ab}\nabla\mu_b,\\
\nonumber
\sigma_{ab}&=&\frac{1}{3T}\int_{t>0} dt~d^3r~\langle {\bm j}_a(0)\cdot \bm{j}_b(t,{\bm r})\rangle\\
\nonumber
D_{ab}&=&\sigma_{ac}\chi^{-1}_{cb},
\end{eqnarray}
where $\mu_a$ is the chemical potential for flavor $a$, and $\chi$ is the susceptibility, or charge fluctuation matrix. The electric conductivity, i.e. the response to a gradient of the charge density for uniform strangeness and baryon number, is
\begin{equation}
D_E=(2D_{uu}+D_{dd}-2D_{ud}-D_{du}-D_{su}+D_{sd})/3. 
\end{equation}
At high temperatures, where $D$ becomes proportional to the unit matrix, $D_E\approx D_{uu}\approx D_{dd}\approx D_{ss}$.\footnote{Although off-diagonal elements in $D_{ab}$ can in principle be retained, they make it significantly more difficult to treat diffusion as a random walk.  Here, we neglect any off-diagonal elements in this matrix; the full treatment is described in \cite{Pratt:2019fbj}.}

\begin{figure}
\centerline{\includegraphics[width=0.4\textwidth]{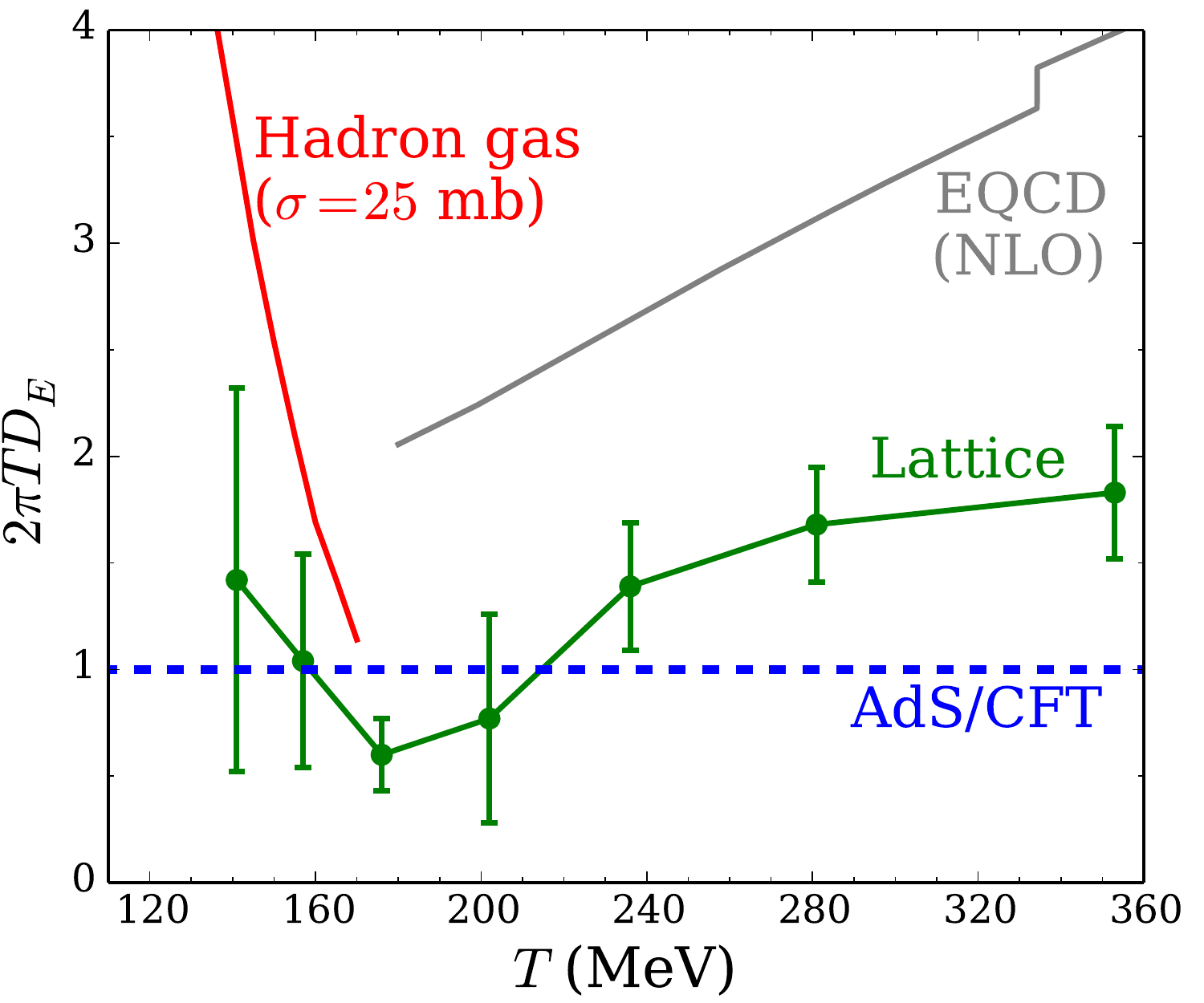}}
\caption{\label{fig:diffcon}(color online) The electric diffusivity, scaled by $2\pi T$, from lattice gauge theory as calculated in \cite{Aarts:2014nba} (green points). A hadron gas with a fixed 25-mb cross section (red line) has significantly higher diffusivity, as does a perturbative approach, EQCD, to next-to-leading order \cite{Ghiglieri:2018dib} (grey line). For AdS/CFT, the value is unity (blue dashes). This collection of calculations illustrates the uncertainty with which the diffusivity is understood.}
\end{figure}

Lattice results for the diffusivity have been obtained \cite{Aarts:2014nba,Amato:2013naa} despite the challenges in extracting transport coefficients from lattice calculations, due to the difficulty in evaluating real-time correlations. For a strongly coupled liquid, AdS/CFT provides a value of  $1/(2\pi T)$ for light quarks \cite{Policastro:2002se}, whereas for heavy quarks the value can approach zero for infinite couplings or colors \cite{CasalderreySolana:2006rq}. The values extracted from lattice are indeed of the order of $1/(2\pi T)$, suggesting that the diffusivity, like the viscosity, is characteristic of strong coupling. For the lattice calculations plotted as a function of $T$, $D$ dips near $T_c$ and falls below the AdS/CFT value (cf. Fig.~\ref{fig:diffcon}).  $D$ can also be calculated perturbatively using an approach known as ``electrostatic QCD" (EQCD) \cite{Ghiglieri:2018dib}.  As shown in Fig.~\ref{fig:diffcon}, this approach applies only above $T_c$ and yields values a few times larger than the lattice results.  Alternatively, one can estimate the conductivity from a quasi-particle picture if one has first an estimate of the relaxation time,
\begin{eqnarray}
\sigma_{ab}&\approx&\int \sum_h d_h\frac{d^3p}{(2\pi)^3}f_h(\bm{p})q_{ha}q_{hb}\frac{|\bm{v}_p|^2}{3}\tau_h(\bm{p}),
\end{eqnarray}
where $\tau_h$ is the relaxation time for a hadron of type $h$, spin degeneracy $d_h$ and charge $q_{ha}$, to lose its correlation with its original velocity. For fixed $s$-wave cross sections, one can estimate the lifetime for a given momentum by calculating the rate at which the correlation, $\langle\bm{v}(0)\bm{v}(t)\rangle$, changes at $t=0$ for each mode ${\bm p}$, then use the inverse rate as the lifetime above. Such a calculation is displayed in Fig. \ref{fig:diffcon} for a hadron gas with 25 mb cross sections, a value consistent with expectations for a hadron gas \cite{Greif:2017byw,Greif:2016skc,Hammelmann:2018ath}. This estimate lies significantly above the lattice prediction. In more realistic pictures, one would include resonant scatterings, and the $s$-wave cross section would be much smaller. But for the purpose of giving an estimate, 25 mb is characteristic of an average cross section in a hadronic system. None of the calculations in Fig. \ref{fig:diffcon} are without significant error and uncertainty, and the four examples are by no means exhaustive in representing the literature, but the variance of these calculations provides a rough picture of how uncertain knowledge is of the diffusivity of high density matter.

Here, a method is proposed for experimentally determining the diffusivity of the light ($u$,$d$,$s$) quarks. Local charge conservation demands that quarks are produced simultaneously with antiquarks, and if one knows the times at which such production occurs, one can constrain the diffusivity  by measuring the correlation of balancing charges in relative momentum. These correlations, which are typically binned in terms of relative rapidity or relative angle, are highly correlated with separation in coordinate space due to the strong collective flow present in these collisions.

In \cite{Pratt:2018ebf} a detailed simulation of the production and diffusion of balancing charges was presented. The evolution of charge correlations was superimposed onto a state-of-the art description of the dynamics, based on hydrodynamics \cite{Shen:2014vra} for higher temperatures and using a hadronic simulation for the hadronic stage and for dissolution. The details of the hydrodynamic evolution were adjusted to reproduce typical multiplicities and $p_T$-spectra, as described in \cite{Pratt:2017oyf,Pratt:2018ebf}. The parameters, however, were not adjusted to represent the best fit to a wide variety of data, and the sensitivity of results to the particular choice of parameters is discussed in the summary. To model the evolution of charge correlations, the charge correlator is first split into local (equilibrated) and non-local (inequilibrated) pieces:
\begin{eqnarray}
C_{ab}(\vec{r}_1,\vec{r}_2,t)
	& \equiv & \langle \rho_a(\vec{r}_1)\rho_b(\vec{r}_2)\rangle \\
	&=&\chi_{ab}(\vec{r}_1,t)\delta(\vec{r}_1-\vec{r}_2)+C'_{ab}(\vec{r}_1,\vec{r}_2,t)
\end{eqnarray}
The non-local piece $C'_{ab}$, which probes the effects of diffusion in an evolving system, is then evolved according to the following diffusion equation:
\begin{eqnarray}
\partial_t C'_{ab}(\vec{r}_1,\vec{r}_2,t)
	&=& D(\nabla_1^2+\nabla_2^2)C_{ab}(\vec{r}_1,\vec{r}_2,t) \nonumber\\
	&& +S_{ab}(\vec{r}_1,t)\delta(\vec{r_1}-\vec{r}_2).
\end{eqnarray}
The source function was determined from the evolution of the susceptibility, assuming the matter maintains local chemical equilibrium \cite{Pratt:2017oyf,Pratt:2018ebf}:
\begin{eqnarray}
S_{ab}(\bm{r},t)&=&(\partial_t+\bm{v}\cdot\nabla+\nabla\cdot\bm{v})\chi_{ab}(\bm{r},t).
\end{eqnarray}
This choice enforces the fact that the correlation $C_{ab}(\vec{r}_1,\vec{r}_2)$ must integrate to zero due to charge conservation, or equivalently that
\begin{eqnarray}
\int d\vec{r}_1~C'_{ab}(\vec{r}_1,\vec{r}_2)&=&\chi_{ab}(\vec{r}_2).
\end{eqnarray}

As described in \cite{Pratt:2018ebf}, the diffusion equation for $C'_{ab}(\vec{r}_1,\vec{r}_2)$ is solved through a Monte Carlo procedure where the correlation is represented by pairs of sample charges. Pairs are created according to the source function above, where the susceptibility at each point in space-time is found as a function of the local temperature according to lattice calculations \cite{Borsanyi:2011sw,Bellwied:2015lba}. The sample charges are propagated according to a random walk as charges move at the speed of light, with their trajectories punctuated by collisions which randomly reorient their direction. The mean free path is chosen to be consistent with the diffusion constant, and given that most test charges experience a half dozen reorientations, this mimics the diffusion equation, with the caveat that the non-causal tails are naturally pruned by requiring the particles to move at the speed of light.

Given that lattice calculations provide only the diffusion for the electric charge, the calculations here assume that the diffusion matrix is diagonal in the QGP phase. Though this should be true at temperatures well above the transition temperature, $D_{ss}$, $D_{uu}$ and $D_{dd}$ should differ once one approaches the transition region, $160<T<200$ MeV. Non-diagonal elements should then appear. It is not clear whether in a full lattice calculation the elements would differ greatly from one another, or whether the presence of off-diagonal elements would significantly change the results. For this reason, without a better picture of the whole diffusivity matrix, it is difficult to ascertain the degree to which results might change.

When the differential charges, $d q_a$, enter the hadron phase they are translated into differential hadron yields, $dN_h$, using thermal arguments \cite{Pratt:2018ebf},
\begin{eqnarray}
dN_h&=&n_h dq_a\chi^{-1}_{ab}Q_{hb},
\end{eqnarray}
where $n_h$ is the equilibrated density of hadrons of species $h$, $Q_{hb}$ is the charge of type $b$ on said species, and $\chi^{-1}$ is the inverse susceptibility.
For a pair of sample charges, each charge, and its decay products, are tracked through a hadron simulation. The resulting products are then used to build the correlation function. In addition to correlations between the products generated by two correlated charges, additional correlations from the hadron phase are generated from the originally uncorrelated particles produced at the interface between the hadronic simulation and the hydrodynamic interface. These come mainly from decays, but are complicated by the rescattering of the decay products, which represents additional diffusive spreading of the balancing charges coming from the decay of the original resonances. This second contribution to the correlation was calculated in a brute force manner, by generating uncorrelated particles, simulating their decay and rescattering, and finally, constructing correlations using the entire ensemble of particles. This procedure includes a large number of uncorrelated particles, which do not contribute to the correlation but do contribute to the noise. To overcome the noise the equivalent of 500,000 hadron events were simulated for each choice for the diffusivity. The two contributions are similar in magnitude, and the latter comprised the bulk of the computational requirements for this analysis.

If chemical equilibrium is established, the susceptibility must jump from $\chi=0$ to its equilibrated value. This contributes a sharp peak to the source function at early times. Once the charges of type $a$ and $b$ are created, they diffuse away from one another according to the diffusivity. If chemical equilibrium were not assumed, one could repeat these calculations with $\chi_{ab}(\vec{r},t)$ chosen to be consistent with whatever alternate chemistry one might wish to explore. For the calculations here, it was assumed that chemical equilibrium was established 0.6 fm/$c$ after the beginning of the collision, at the time the hydrodynamic calculation begins. At this time the balancing correlation $C'_{ab}(\vec{r}_1,\vec{r}_2)$ is established with a finite width in relative rapidity as described in \cite{Pratt:2018ebf}. The initial spread of the correlation was given a Gaussian form with a width of $\sigma_0=0.5$ units of spatial rapidity. Even though less than 1 fm/$c$ has transpired at this time, this longitudinal spread is significant due to the large longitudinal velocity gradients at early times. This affects the width of the balance function when binned by rapidity, and in \cite{Pratt:2018ebf} the sensitivity to $\sigma_0$ was analyzed. When viewing the balance function binned in rapidity, it is difficulty to discern whether a broader balance function was due to a larger initial spread, or a larger diffusivity. In contrast, there should be little transverse flow at such early times, at least for central collisions. Thus, the width of the balance function in relative angle should by comparison be much less sensitive to the choice of the starting time or any initial spread of $C'$, meaning that balance functions binned by relative angle should provide a cleaner signal from which to infer the diffusivity.

Correlations are manifested as charge balance functions,
\begin{eqnarray}
\label{eq:bdef}
B(\Delta \phi)&\equiv&\left\langle N_{+-}(\Delta \phi)+N_{-+}(\Delta \phi)\right.\\
\nonumber
& &\left.-N_{++}(\Delta \phi)-N_{--}(\Delta \phi)\right\rangle/(N_++N_-).
\end{eqnarray}
Here, $N_{qq'}(\Delta \phi)$ denotes the sum over all pairs of charges $qq'$ separated by azimuthal angle $\Delta\phi$, $N_{q}$ is the number of charges of type $q$, and the average covers all events of a given centrality class. Charge balance functions have been measured as a function of relative rapidity, pseudo-rapidity and azimuthal angle, and the charges restricted to specific hadron species. Data from both RHIC and from the LHC have been analyzed \cite{Wang:2012jua,Abelev:2010ab,Adams:2003kg,Aggarwal:2010ya,Abelev:2013csa,Alt:2007hk,Adamczyk:2015yga,Adamczyk:2013hsi,Abelev:2009ac}.

Not surprisingly, $B_{K^+K^-}$ focuses on the correlation between strange quarks \cite{Pratt:2017oyf}. The source function for strangeness, $S_{ss}$, is dominated by the first surge of charge production as the system is initially equilibrated in the first $\lesssim 1$ fm/$c$ of the collision. During the evolution of an idealized QGP of massless quarks and gluons, entropy conservation maintains the number of quarks and $S_{ab}$ vanishes. Once hadrons form the source function again becomes strong, both because hadrons carry multiple quarks and because, by entropy conservation, the number of hadrons roughly equals the number of quarks in the QGP \cite{Bass:2000az,Pratt:2012dz}. In contrast to the source functions for up and down quarks, the source function for strangeness remains small during hadronization \cite{Pratt:2017oyf,Borsanyi:2011sw} due to the larger mass of strange hadrons, which suppresses the production of strange quarks once the hadron phase is realized. The one exception to this comes from the decay of the phi meson, $\phi\rightarrow K^+K^-$, but given that the phi peak is narrow, $4.3$ MeV, their contribution can be separated from the analysis to better focus on the contributions to the correlation from charge generated during the QGP stage. Indeed, STAR analyses of kaon balance functions features a subtraction of the $\phi$ meson, and contributions from $\phi$ mesons are also subtracted from the numerator of the balance functions calculated in this study. Even though up and down quarks are copiously created during hadronization, the effective source function for baryon number is small during hadronization, and even becomes negative below $T_c$ \cite{Pratt:2017oyf}, due to baryon annihilation in the hadronic stage, which is driven by the high mass of baryons. Thus, the $K^+K^-$ and $p\bar{p}$ balance functions should be more sensitive to the diffusivity because the source functions that drive them are concentrated at early times, allowing diffusion to play the main role in their separation.

\begin{figure*}
\centering
\includegraphics[width=0.4\textwidth]{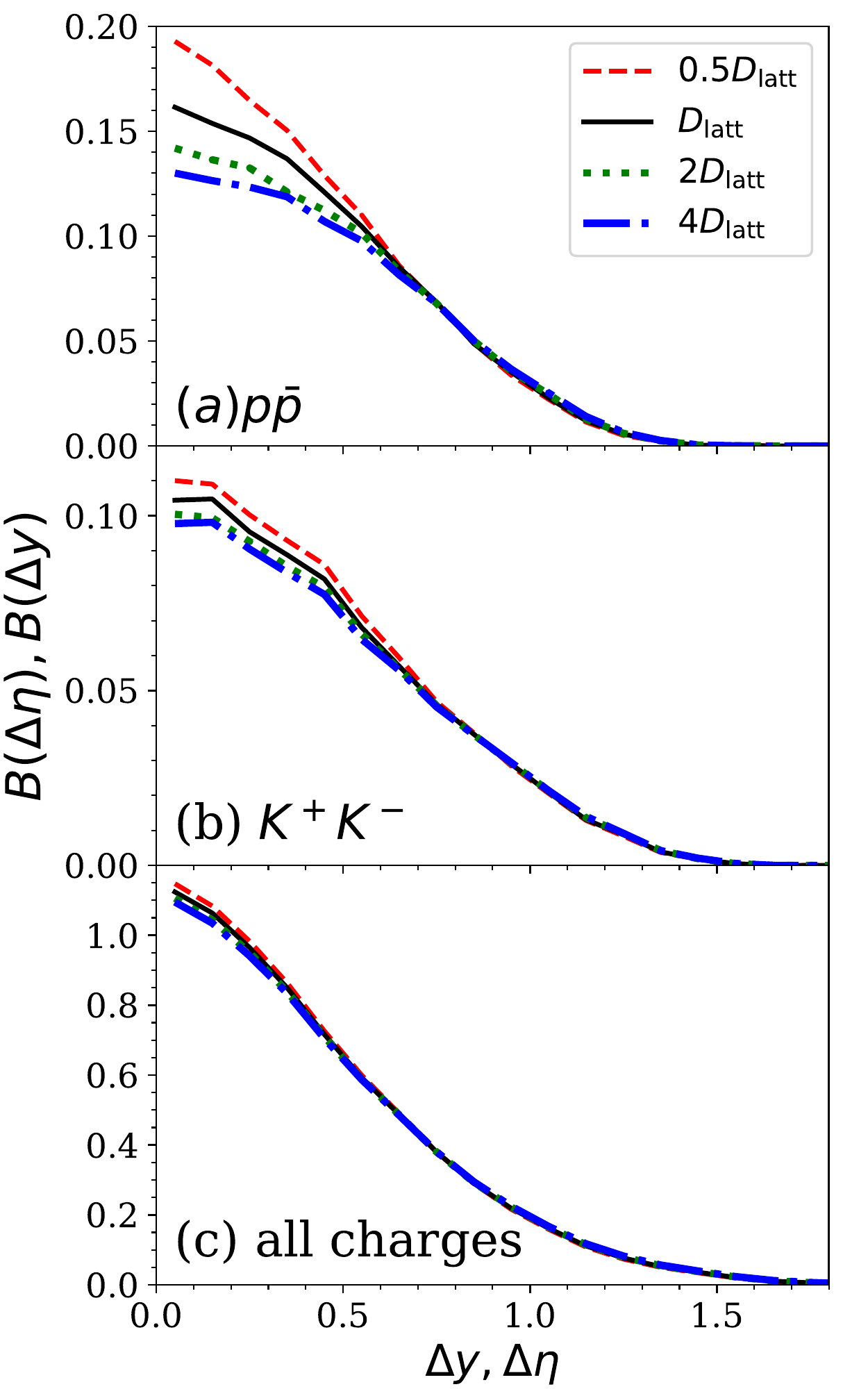}\hspace*{0.1\textwidth}\includegraphics[width=0.4\textwidth]{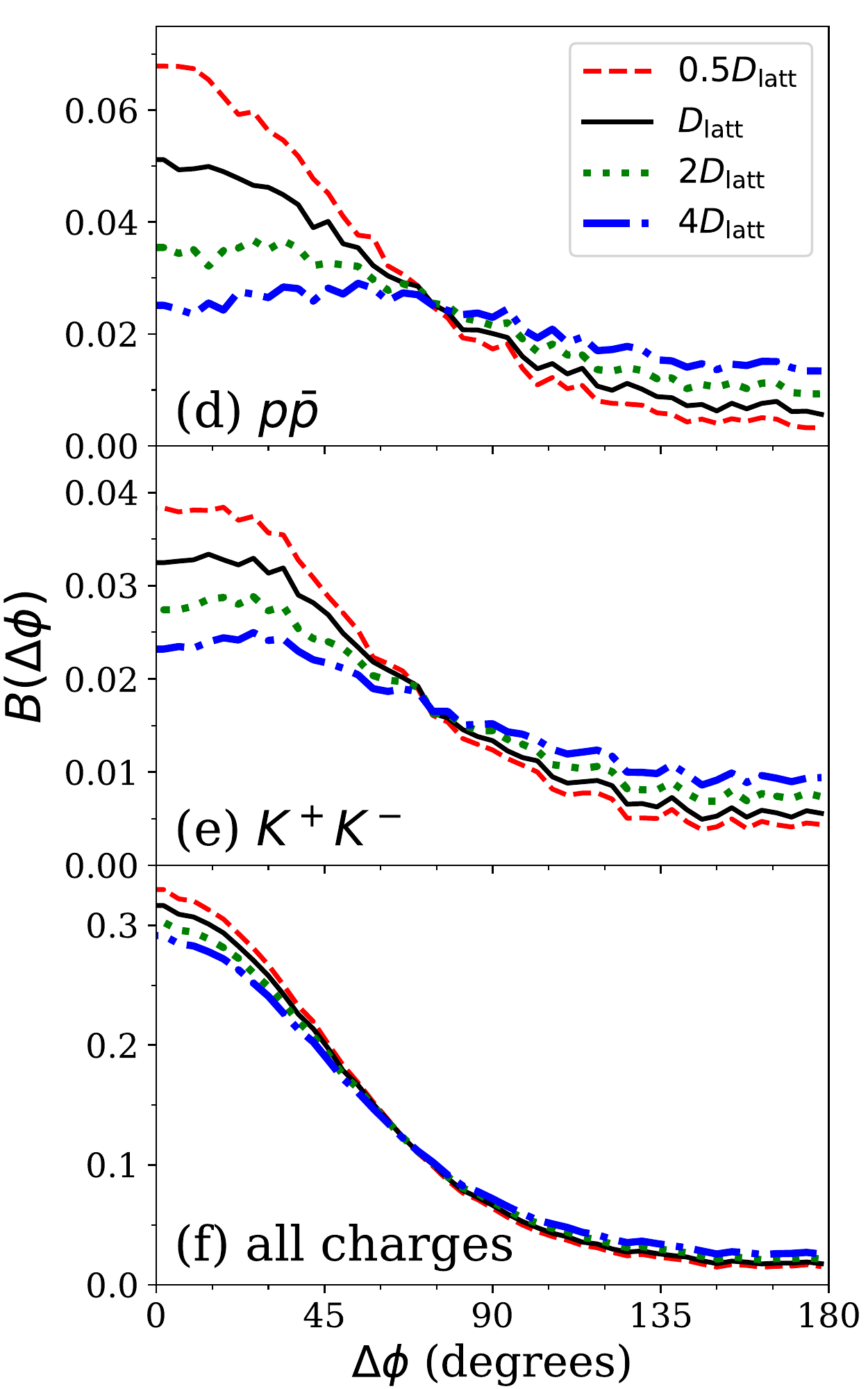}
\caption{\label{fig:bf_phi}(color online)
Model calculations of charge balance functions binned by relative rapidity (panels a-c) or by relative azimuthal angle (d-f) for four choices of the diffusion constant. Balance functions created using only $p,\bar{p}$ (a) or using only $K^+,K^-$ (b) are binned by rapidity $\Delta y$, whereas results using all charged particles (c) are binned by pseudo-rapidity $\Delta \eta$. Calculations using the diffusivity from lattice calculations \cite{Aarts:2014nba} are represented by black lines, while using half (red dashes), double (green dotted) and quadruple (blue dot-dashed) the lattice results illustrate how the balance function broadens for increasing diffusivity. Results for $p\bar{p}$ (a,d) and $K^+K^-$ (b,e) are especially sensitive to diffusion because the source terms for the baryon-baryon and strange-strange correlation functions are concentrated on early times, which gives diffusion more opportunity to separate balancing charges. The balance functions in the left-side panels, which are binned by rapidity, are also significantly sensitive to how charges are separated at the initial time when hydrodynamics is instantiated, or to the equivalently the time at which chemical equilibrium occurs due to the large initial longitudinal flow. In contrast, in the right-side panels, which are binned by relative angle, results are less sensitive to the specifics of the physics before one fm/$c$ because transverse flow requires time to develop.}
\end{figure*}
Figure \ref{fig:bf_phi} shows balance functions for three cases: all positive/negative particles, $K^+K^-$ and $p\bar{p}$. The methods are the same as described above and applied in \cite{Pratt:2018ebf}. In the hydrodynamic stage, during which the matter is largely in the QGP phase, the charge-charge correlations were evolved according to four different choices for the diffusivity. First, they were evolved according to $D(T)$ reported from lattice calculations \cite{Aarts:2014nba}, exactly as in \cite{Pratt:2018ebf}. Then, the calculations were repeated with half that value, double that value, and finally, four times the lattice diffusivity. The analysis was restricted to very central events, 0-5\% centrality. In each case the balance functions are broader for the larger diffusivities. The balance function for all charges is least sensitive because it is dominated by later-stage production of charge associated with hadronization.  In contrast, the $K^+K^-$ and $p\bar{p}$ balance functions broaden significantly. Unfortunately, experimental results for $K^+K^-$ and $p\bar{p}$ balance functions have only been reported binned by relative rapidity thus far. Preliminary results for all charges have been reported by STAR \cite{Wang:2012jua}, but are marred by the effects of experimental sector boundaries.  

The results of Fig. \ref{fig:bf_phi} suggest that both $K^+K^-$ and $p\bar{p}$ are promising for constraining the diffusivity of the QGP. This was expected, given that the source functions driving the those balance functions were concentrated at early times. However, the $p\bar{p}$ results are strongly sensitive to the choice of the QGP to hadron phase transition temperature. Because of the large baryon mass, the equilibrium number of baryons falls rapidly with falling temperature once one enters the hadronic phase, which corresponds to the introduction of negative source functions. Equivalently, in the hadron stage the effects of baryon annihilation can significantly alter the shape of the $p\bar{p}$ balance function, leading to a dip of the balance function at small relative angle, as well as (due to normalization constraints) an accompanying increase at large relative angle. A careful analysis of annihilation effects requires consistently accounting for regeneration \cite{Pan:2014caa,Steinheimer:2017vju,Steinheimer:2012rd}, and until such a consistent analysis is performed, one must remain cautious in interpreting $p\bar{p}$ balance function results.

To most clearly constrain the diffusivity, it is better to focus on balance functions binned by relative angle. As mentioned above,  balancing charges produced in the first 1 fm$/c$ might separate significantly along the beam direction by the time the hydrodynamic description is instantiated. Due to the large velocity gradient along the beam axis at early times, $dv_z/dz\approx 1/\tau$, a separation of $1/2$ fm at a time $\tau=1/2$ fm/$c$ translates to a separation of an entire unit of spatial rapidity. Disentangling the longitudinal separations related to pre-equilibrium dynamics from the effects of diffusion could therefore be problematic. Because there are no large transverse velocity gradients at early times, the transverse separation should be dominated by the effects of diffusion, especially for the large sources in central collisions.

Of the several balance functions, the $K^+K^-$ balance function binned by relative angle holds the most promise for robustly constraining the diffusivity of the QGP. Because the source functions that contribute to the $K^+K^-$ balance functions are dominated by the initial thermalization of the QGP, the spread of the balance function in relative angle will be dominated by diffusion. Nonetheless uncertainties remain. First, there are contributions from later times through $\phi$ meson decays. The $\phi$ contribution does not affect the number of balancing pairs at relative large angle, but it does affect the normalization. Fortunately, even if this affects the normalization by several tens of percent, careful measurement of $\phi$ mesons can ensure that this is taken into account so that the normalizations are uncertain at the five percent level. Viscous effects can increase entropy at the ten percent level, and should be accompanied by a corresponding increase in multiplicity, including the multiplicity of strange-antistrange quark pairs. If viscous effects are understood at a factor-of-two level, and if the effects are $\lesssim 10$\%, this also represents roughly a five percent uncertainty to the balance function. Finally, the calculations here assume that chemical equilibrium has taken hold by one fm/$c$ into the collision. If equilibrium, and the corresponding production of strange quarks, were to take hold at later times, the charge balance function would narrow. The best way to test whether quark production takes place early is to analyze the balance functions, especially the $p\bar{p}$ and $K^+K^-$ balance functions, in terms of relative rapidity. Balancing charges that are separated by more than a unit of rapidity are indicative of early production. Current experiments at the LHC and at RHIC have difficulty measuring identified pairs separated by $\Delta y>1$ due to the limited acceptance, but given that the normalization of the balance functions are constrained by chemistry and charge conservation, one can infer the overall strength of the balance functions outside the acceptance. Comparison with simple parameterized thermal models have made it clear that there is significant early production. The fact that the $K^+K^-$ balance function is significantly broader than the $\pi^+\pi^-$ balance functions \cite{Pratt:2015jsa,Wang:2012jua}, and that the width of the $K^+K^-$ balance function, unlike the width of the $\pi^+\pi^-$ balance function does not narrow with increasing centrality, make it clear that strange quark production is early. However, quantifying exactly how early can be difficult. For a central collision, the width of the $K^+K^-$ balance functions in relative rapidity depend sensitively on whether the quarks are produced at 0.5 or 1.5 fm/$c$ and can represent a factor of two difference in the spatial separation of quarks along the beam axis. However, for transverse separation, a difference of one fm/$c$ out of the $\sim 5$ fm/$c$ duration of the QGP phase should affect the separation at the 10\% level. Thus, if careful analysis of all the balance functions, binned in both relative rapidity and relative angle can pinpoint the initial thermalization to $\tau\lesssim 1$ fm/$c$, the uncertainties associated with the creation time should be at the ten percent level.

\begin{figure}
\centerline{\includegraphics[width=0.45\textwidth]{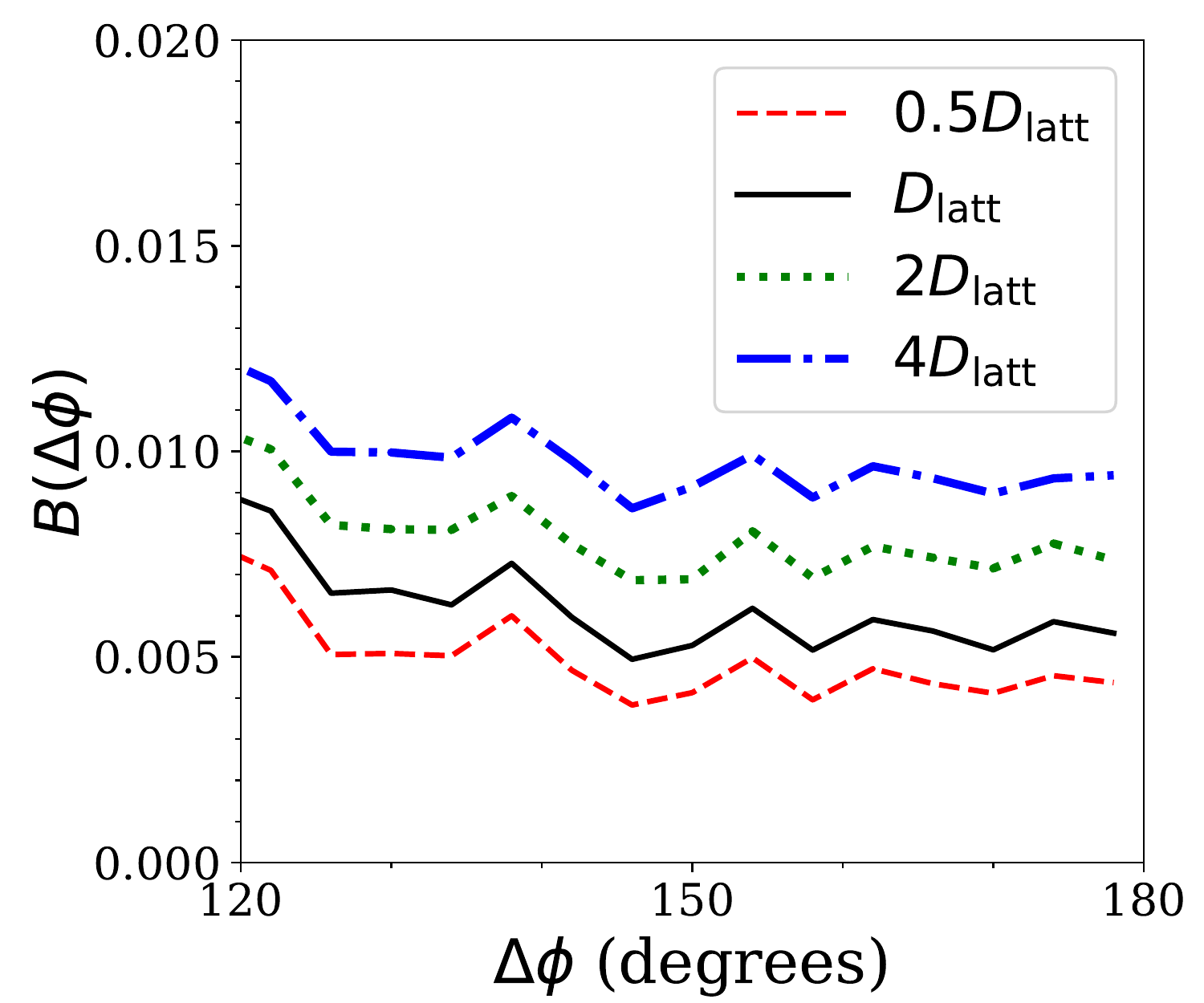}}
\caption{\label{fig:blowup}
The tail of the $K^+K^-$ charge balance function binned by relative momentum from Fig. \ref{fig:bf_phi} is magnified to better illustrate the sensitivity to the diffusivity. If the various sources of uncertainty that affect the normalization of the balance function can be reduced to the 15\% level, this should enable constraint of the diffusivity to better than a factor of two.}
\end{figure}
The sum of the uncertainties could therefore realistically attain the 15\% level if the various complementary considerations listed above were all analyzed. Figure \ref{fig:blowup} expands the tail of the $K^+K^-$ balance functions from Fig. \ref{fig:bf_phi}. The height of the balance function varies by a factor of two when varying the diffusivity from one half to four times the lattice value. If the height of the experimental balance functions were changed by 15\%, the extracted diffusivity would be modified by roughly 50\%. Thus, a combination of improved experimental statistics and careful theoretical analysis might constrain the diffusivity of strange quarks to better than a factor of two. This resolution would be similar to how well the shear viscosity, another fundamental transport coefficient of the QGP, has been extracted from experiment.  

\begin{acknowledgments}
This work was supported by the Department of Energy Office of Science through grant number DE-FG02-03ER41259 and through grant number DE-FG02-87ER40328.  C.~P.~ is funded by the CLASH project (KAW 2017-0036), and gratefully acknowledges the use of computing resources from both the Minnesota Supercomputing Institute
(MSI) at the University of Minnesota and the Ohio Supercomputer Center
\cite{OhioSupercomputerCenter1987}.
\end{acknowledgments}


\begin{thebibliography}{99}
\bibitem{Pratt:2015zsa} 
  S.~Pratt, E.~Sangaline, P.~Sorensen and H.~Wang,
  Phys.\ Rev.\ Lett.\  {\bf 114}, 202301 (2015).
  
	
\bibitem{Pratt:2015jsa} 
  S.~Pratt, W.~P.~McCormack and C.~Ratti,
  Phys.\ Rev.\ C {\bf 92}, 064905 (2015).
  
\bibitem{Bernhard:2016tnd} 
  J.~E.~Bernhard, J.~S.~Moreland, S.~A.~Bass, J.~Liu and U.~Heinz,
  Phys.\ Rev.\ C {\bf 94}, no. 2, 024907 (2016).
  
\bibitem{Bernhard:2015hxa} 
  J.~E.~Bernhard, P.~W.~Marcy, C.~E.~Coleman-Smith, S.~Huzurbazar, R.~L.~Wolpert and S.~A.~Bass,
  Phys.\ Rev.\ C {\bf 91}, no. 5, 054910 (2015).

\bibitem{Burke:2013yra}
  K.~M.~Burke {\it et al.} [JET Collaboration],
  Phys.\ Rev.\ C {\bf 90}, no. 1, 014909 (2014).

\bibitem{He:2018gks} 
  Y.~He, L.~G.~Pang and X.~N.~Wang,
  arXiv:1808.05310 [hep-ph].

\bibitem{Xu:2017obm} 
  Y.~Xu, J.~E.~Bernhard, S.~A.~Bass, M.~Nahrgang and S.~Cao,
  Phys.\ Rev.\ C {\bf 97}, no. 1, 014907 (2018).

\bibitem{Borsanyi:2011sw}
  S.~Borsanyi, Z.~Fodor, S.~D.~Katz, S.~Krieg, C.~Ratti and K.~Szabo,
  JHEP {\bf 1201}, 138 (2012).
  
\bibitem{Bellwied:2015lba}
R.~Bellwied, S.~Borsanyi, Z.~Fodor, S.~D.~Katz, A.~Pasztor, C.~Ratti and K.~K.~Szabo,
Phys. Rev. D \textbf{92}, no.11, 114505 (2015).

\bibitem{Greif:2017byw} 
  M.~Greif, J.~A.~Fotakis, G.~S.~Denicol and C.~Greiner,
  Phys.\ Rev.\ Lett.\  {\bf 120}, 242301 (2018).

\bibitem{Pratt:2019fbj} 
  S.~Pratt,
  arXiv:1908.01053 [nucl-th].

\bibitem{Aarts:2014nba} 
  G.~Aarts, C.~Allton, A.~Amato, P.~Giudice, S.~Hands and J.~I.~Skullerud,
  JHEP {\bf 1502}, 186 (2015).

\bibitem{Amato:2013naa} 
  A.~Amato, G.~Aarts, C.~Allton, P.~Giudice, S.~Hands and J.~I.~Skullerud,
  Phys.\ Rev.\ Lett.\  {\bf 111}, no. 17, 172001 (2013).

\bibitem{Policastro:2002se} 
  G.~Policastro, D.~T.~Son and A.~O.~Starinets,
  JHEP {\bf 0209}, 043 (2002).

\bibitem{CasalderreySolana:2006rq} 
  J.~Casalderrey-Solana and D.~Teaney,
  Phys.\ Rev.\ D {\bf 74}, 085012 (2006).

\bibitem{Ghiglieri:2018dib} 
  J.~Ghiglieri, G.~D.~Moore and D.~Teaney,
  JHEP {\bf 1803}, 179 (2018).
  
  
\bibitem{Greif:2016skc} 
  M.~Greif, C.~Greiner and G.~S.~Denicol,
  Phys.\ Rev.\ D {\bf 93}, no. 9, 096012 (2016)
  Erratum: [Phys.\ Rev.\ D {\bf 96}, no. 5, 059902 (2017)]
  
\bibitem{Hammelmann:2018ath} 
  J.~Hammelmann, J.~M.~Torres-Rincon, J.~B.~Rose, M.~Greif and H.~Elfner,
  Phys.\ Rev.\ D {\bf 99}, no. 7, 076015 (2019).

\bibitem{Pratt:2018ebf} 
  S.~Pratt and C.~Plumberg,
  to appear in Phys. Rev. C, arXiv:1812.05649 [nucl-th].

\bibitem{Shen:2014vra} 
  C.~Shen, Z.~Qiu, H.~Song, J.~Bernhard, S.~Bass and U.~Heinz,
  Comput.\ Phys.\ Commun.\  {\bf 199}, 61 (2016)

\bibitem{Pratt:2017oyf}
  S.~Pratt, J.~Kim and C.~Plumberg,
  Phys.\ Rev.\ C {\bf 98}, no. 1, 014904 (2018).

\bibitem{Wang:2012jua} 
  H.~Wang, Ph.D. Thesis,
  arXiv:1304.2073 [nucl-ex].

\bibitem{Abelev:2010ab} 
  B.~I.~Abelev {\it et al.} [STAR Collaboration],
  Phys.\ Lett.\ B {\bf 690}, 239 (2010).

  N.~Li {\it et al.} [STAR Collaboration],

\bibitem{Adams:2003kg} 
  J.~Adams {\it et al.} [STAR Collaboration],
  Phys.\ Rev.\ Lett.\  {\bf 90}, 172301 (2003).
   
\bibitem{Aggarwal:2010ya} 
  M.~M.~Aggarwal {\it et al.} [STAR Collaboration],
  Phys.\ Rev.\ C {\bf 82}, 024905 (2010).


\bibitem{Abelev:2013csa}
  B.~Abelev {\it et al.} [ALICE Collaboration],
  Phys.\ Lett.\ B {\bf 723}, 267 (2013).
  
\bibitem{Alt:2007hk} 
  C.~Alt {\it et al.} [NA49 Collaboration],
  Phys.\ Rev.\ C {\bf 76}, 024914 (2007).

\bibitem{Adamczyk:2015yga} 
  L.~Adamczyk {\it et al.} [STAR Collaboration],
  Phys.\ Rev.\ C {\bf 94}, no. 2, 024909 (2016).

\bibitem{Adamczyk:2013hsi} 
  L.~Adamczyk {\it et al.} [STAR Collaboration],
  Phys.\ Rev.\ C {\bf 88}, no. 6, 064911 (2013).

\bibitem{Abelev:2009ac} 
  B.~I.~Abelev {\it et al.} [STAR Collaboration],
  Phys.\ Rev.\ Lett.\  {\bf 103}, 251601 (2009).

\bibitem{Bass:2000az} 
  S.~A.~Bass, P.~Danielewicz and S.~Pratt,
  Phys.\ Rev.\ Lett.\  {\bf 85}, 2689 (2000).

\bibitem{Pratt:2012dz} 
  S.~Pratt,
  Phys.\ Rev.\ Lett.\  {\bf 108}, 212301 (2012).


\bibitem{Pan:2014caa} 
  Y.~Pan and S.~Pratt,
  Phys.\ Rev.\ C {\bf 89}, no. 4, 044911 (2014).

\bibitem{Steinheimer:2017vju} 
  J.~Steinheimer, J.~Aichelin, M.~Bleicher and H.~Stöcker,
  Phys.\ Rev.\ C {\bf 95}, no. 6, 064902 (2017).
 
\bibitem{Steinheimer:2012rd} 
  J.~Steinheimer, J.~Aichelin and M.~Bleicher,
  Phys.\ Rev.\ Lett.\  {\bf 110}, no. 4, 042501 (2013).
  
  \bibitem{OhioSupercomputerCenter1987}
Ohio~Supercomputer Center.
\newblock Ohio supercomputer center, 1987.

\end{thebibliography}
\end{document}